\documentclass[preprintnumbers,10pt,nofootinbib]{revtex4}

\usepackage{amsmath,latexsym,amssymb,amsfonts}
\usepackage[dvips]{graphicx,color}
\usepackage{bm}

\begin{document}


\title{\textbf{\textit{Annihilation-to-nothing}: DeWitt boundary condition inside a black hole}}

\author{
Suddhasattwa Brahma$^{a}$\footnote{{\tt suddhasattwa.brahma@gmail.com}},
Che-Yu Chen$^{b}$\footnote{{\tt b97202056@gmail.com}},
and
Dong-han Yeom$^{c,d}$\footnote{{\tt innocent.yeom@gmail.com}}
}

\affiliation{
$^{a}$Department of Physics, McGill University, Montr\'eal, QC H3A 2T8, Canada\\
$^{b}$Institute of Physics, Academia Sinica, Taipei 11529, Taiwan\\
$^{c}$Department of Physics Education, Pusan National University, Busan 46241, Republic of Korea\\
$^{d}$Research Center for Dielectric and Advanced Matter Physics, Pusan National University, Busan 46241, Republic of Korea
}

\begin{abstract}
In canonical quantum gravity, the wave function for a hypersurface inside a Schwarzschild black hole can be obtained by solving the Wheeler-DeWitt equation. What is of prime importance is the behavior of the wave function for the future boundary near the singularity, and the DeWitt boundary condition implies that it should vanish here. In this paper, we provide several generalizations, and new interpretations, of the DeWitt boundary condition. First, we summarize existing works on the wave function inside the black hole to justify the DeWitt boundary condition. Next, we investigate the wave function for the collapsing null shell to show that due to the reflection symmetry in space and time, there exists a destructive interference near the singularity and hence a vanishing boundary condition can be natural. If we extend this point of view to the black hole spacetime itself, then the DeWitt boundary condition is equivalent to saying that there exists a symmetric anti-black hole contribution, such that eventually these two geometries are \textit{annihilated-to-nothing} near the quantum transition surface. This symmetric model can be realized within black hole models of loop quantum gravity with a novel interpretation for the arrow(s) of time.
\end{abstract}

\maketitle

\newpage

\tableofcontents

\newpage

\section{Introduction}
Black holes have been a driving force behind developing quantum theories of gravity. This is particularly so since understanding the nature of the singularity inside a black hole is an outstanding central problem of quantum gravity, and indeed, in modern physics \cite{Hawking:1970zqf}. It is also closely related to resolving the information loss problem of black holes \cite{Hawking:1976ra}.  

One of the most conservative ways to deal with the black hole singularity is to canonically quantize the internal spacetime of a black hole \cite{DeWitt:1967yk}. On imposing a specific metric ansatz, one can derive and solve the quantum Hamiltonian constraint equation, or the so-called Wheeler-DeWitt equation \cite{Cavaglia:1994yc}. The problem becomes tractable when considering only the interior of the Schwarzschild black hole, which can be assumed to be described by an anisotropic, and yet homogeneous, metric \cite{Kantowski:1966te}. In this case, there remain only two dynamical (metric) degrees of freedom and the Wheeler-DeWitt equation reduces to a partial differential equation with two canonical variables. Homogeneity plays the crucial role of reducing it to a manageable minisuperspace model.

Although solving the Wheeler-DeWitt equation for full quantum gravity poses a tremendously difficult problem, minisuperspace models have been met with much more considerable success, in both cosmology \cite{Vilenkin:1987kf} and black hole spacetimes \cite{Kuchar:1994zk}, due to their symmetry-reduced nature. In order to find solutions for the Wheeler-DeWitt equation, one needs to consider the boundary condition(s) of the wave function. One of the natural boundary conditions is to assume that the wave function must behave as classical at the event horizon. In other words, the steepest-descent of the wave function should coincide with the classical solution \cite{Bouhmadi-Lopez:2019kkt,Yeom:2019csm}. This condition is synonymous with the physical requirement that for macroscopic black holes, one does not expect any large quantum gravity effects to appear at the horizon. As soon as we impose this condition, as well as assume the bounded nature of the solution, the wave function is uniquely defined and the steepest-descent follows the classical solution down to the singularity. This, in turn, means that assuming such a reasonable boundary condition as above would imply that the Wheeler-DeWitt quantization fails to resolve the classical black hole singularity.

The question, therefore, is as follows --- Where can one expect to see quantum gravitational effects? A way to avoid the above conclusion would simply be to accept the failure of simple Wheeler-DeWitt quantization and assume that one needs further inputs regarding the quantum nature of the underlying spacetime in order to find such effects. However, before resigning ourselves to such a drastic eventuality, let us note that there have been a series of recent developments within the Wheeler-DeWitt picture which offer an alternative resolution.

Firstly, it was shown that there exists a quantum ``bounce'' near $r \sim M$ hypersurface, where $r$ denotes the usual radius and $M$ is the black hole mass \cite{Bouhmadi-Lopez:2019kkt,Yeom:2019csm}. In addition, there was another  proposal such that the wave function should vanish at the singularity $r \sim 0$ itself \cite{Perry:2021mch}. The takeaway message from these papers is that it is possible to have an annihilation of the wave function on specific hypersurfaces with suitable assumptions. If these proposals are correct, then it is reasonable to think that imposing a vanishing boundary condition at the singularity, or the so-called DeWitt boundary condition \cite{DeWitt:1967yk}, is a generic nature of the Wheeler-DeWitt wave function. To be clear, the claim is that the wave function solution of the Wheeler-DeWitt equation vanishes \textit{naturally} at the singularity. However, do we have any strong argument or evidence that this is generic, when considering realistic gravitational collapses or within a fundamental approach to quantum gravity? 

In this paper, we would like to answer these questions. First, in Sec.~\ref{sec:dewitt}, we consider the thin-shell collapse and quantize the shell dynamics using the approach of Hajicek and Kiefer \cite{Hajicek:2000mh}. The DeWitt boundary condition for collapsing matter can be shown to follow explicitly, although we need to revisit the interpretation. Second, in Sec.~\ref{sec:black}, we consider the possibility of having a vanishing boundary condition near the singularity in the context of loop quantum gravity inspired models. Recently, several models have been proposed in which there is a big bounce near the putative singularity due to the loop quantum gravity corrections \cite{Ashtekar:2018lag,Bodendorfer:2019xbp}. It is shown that these models can have a new interpretation in light of the DeWitt boundary condition. Finally, we speculate how the annihilation of the wave function can be made compatible with dynamical signature-change in loop quantum gravity \cite{Sig_Change}, while invoking a final-state like condition for black holes \cite{Horowitz:2003he}. But before that, in the next section, we will first begin with a lightening review of previous literature and the intricacies of the DeWitt boundary condition for the vacuum Schwarzschild spacetime.

\section{Preliminary: wave function inside a vacuum Schwarzschild black hole}
We first show how the wave function can get annihilated inside a vacuum black hole spacetime due to destructive interference of the wave function. In particular, we show that on including extra ingredients to the Wheeler-DeWitt equation as one approaches the singularity, it is possible to recover the DeWitt boundary condition within such a formulation.

\subsection{Steepest-descent up to the singularity}
In this subsection, we briefly review the picture of vanishing wave functions inside black holes, as first described in \cite{Bouhmadi-Lopez:2019kkt,Yeom:2019csm}. This picture is based on some particular solutions of the Wheeler-DeWitt equation such that the steepest-descent of the wave function follows the classical trajectory at the horizon. We choose the Kantowski-Sachs anisotropic metric ansatz for the spacetime inside the horizon as follows \cite{Kantowski:1966te}:
\begin{eqnarray}\label{KS}
ds^{2} = - N^{2}(t) dt^{2} + a^{2}(t) dr^{2} + \frac{r_{s}^{2} b^{2}(t)}{a^{2}(t)} d\Omega^{2}\,,
\end{eqnarray}
where $N(t)$ is the lapse function, $r_{s}$ is the Schwarzschild radius, and $a(t)$ and $b(t)$ denote two canonical variables. Here, the classical Schwarzschild solution corresponds to the following relation:
\begin{eqnarray}
\frac{1}{b} = a + \frac{1}{a}\,,
\end{eqnarray}
where this relation is invariant with respect to the choice of the lapse function $N(t)$. From this metric ansatz, one can derive the Wheeler-DeWitt equation \cite{Cavaglia:1994yc}:
\begin{eqnarray}
\left( \frac{\partial^{2}}{\partial X^{2}} - \frac{\partial^{2}}{\partial Y^{2}} + 4 r_{s}^{2} e^{2Y} \right) \Psi \left(X,Y\right) = 0\,,
\end{eqnarray}
where $X = \ln a$ and $Y = \ln b$.

In order to solve the partial differential equation, one first introduces the usual separation of variables, say $\Psi = \phi(X) \psi(Y)$, for which the Wheeler-DeWitt equation takes the form:
\begin{eqnarray}
\left(\frac{d^{2}}{dX^{2}} + k^{2}\right) \phi(X) &=& 0\,, \\
\left(\frac{d^{2}}{dY^{2}} - 4 r_{s}^{2} e^{2Y} + k^{2}\right) \psi(Y) &=& 0\,,
\end{eqnarray}
where $k^{2}$ is the separation constant. Interestingly, this set of equations is equivalent to the two-dimensional Schr\"odinger equation of quantum mechanics, where $X$ is a time-like direction and $Y$ is a space-like direction with the potential barrier $\sim e^{2Y}$.

As is evident from above, there is only a potential barrier along the $Y$-direction, and it is divergent as $Y$ goes to infinity. Hence, if there is an incoming mode along the $+Y$-direction, then there must be an outgoing mode along the $-Y$-direction (unless there exist divergent contributions in the $Y > 0$ region). The classical observer will follow the Ehrenfest theorem, \textit{i.e.}, the peak of the wave function should correspond to the classical solution. Therefore, at the horizon, we need to impose the boundary condition that the wave function has a peak at the classical solution. By imposing these properties, and without loss of generality, we obtain the following form of the solution \cite{Bouhmadi-Lopez:2019kkt}:
\begin{eqnarray}
\Psi(X,Y) = \int_{-\infty}^{\infty} f(k)\, e^{-ikX}\, K_{ik}\,\left(2r_{s} e^{Y}\right)\, dk\,,
\end{eqnarray}
where $K_{ik}$ is the hyperbolic Bessel function. If we choose
\begin{eqnarray}
f(k) = \frac{2A e^{-\sigma^{2}k^{2}/2}}{\Gamma(-ik)r_{s}^{ik}}\,,
\end{eqnarray}
then the classical solution is located at the Gaussian peak of the wave function at the event horizon ($X, Y \rightarrow - \infty$), where $\sigma$ is the standard deviation and $A$ is the normalization constant.

This solution implies that the steepest-descent of the wave function (the ridge of the wave function) follows the classical trajectory. However, since there is a potential barrier along the $Y$-direction, there exists a quantum bounce around $r = r_{s}b/a \sim M$. At this surface, it is possible to impose the DeWitt-like boundary condition inside the black hole. What we mean by the ``DeWitt-like boundary condition'' is that we find a wave function solution that has a hypersurface on which the wave function annihilates. Of course, unlike in the original proposal by DeWitt, this is not the $r=0$ hypersurface, and therefore, one does not find that the wave function annihilates at the singularity. Also, it is worthwhile to mention that this is the \textit{future} boundary condition; hence, we do not prepare this boundary condition at the past hypersurface, but this is a result by solving the Wheeler-DeWitt equation.

Let us end this section with two take-home messages. Firstly, we show that in a non-perturbative quantization scheme of the vacuum Schwarzschild spacetime following the Wheeler-DeWitt equation, it is indeed possible to find quantum effects due to which one cannot faithfully follow the peak of the wave function down to the singularity. This demonstrates that the quantum solution can lead to nice surprises and does not necessarily imply the existence of a singularity as in the classical case, even on imposing a classical boundary condition at the event horizon. In particular, we find that there is a hypersurface $r \sim M$ where the wave function goes to zero. Around the annihilation surface, there appears two pieces of classical branches of the wave function; for each branch, one can impose an arrow of time. Interpreting the arrows of time, one finds two wave packets -- one originating from the event horizon and the other from the singularity -- which destructively interferes at the $r \sim M$ hypersurface. This was termed as the \textit{annihilation-to-nothing} picture in \cite{Bouhmadi-Lopez:2019kkt}. However, what is also apparent from the preceding discussion is that one cannot impose the classical DeWitt condition to have the wave function go to zero at the singularity, once classicality is imposed at the horizon, unless some additional ingredients are invoked. In the next subsection, we show how one can impose the original DeWitt boundary condition within this approach once the Belinsky-Khalatnikov-Lifshitz (BKL) is taken into consideration.

\subsection{BKL conjecture and the DeWitt boundary condition}
If we do not take into account the quantum bouncing hypersurface, the steepest-descent of the wave function will approach the singularity, just as in the classical solution. As mentioned earlier, it would then seem that the boundary condition of imposing classicality at the horizon is going to be incompatible with the vanishing of the wave function at the singularity. However, the way out of this puzzle is in the way of the following realization: As the solution approaches the singularity, the anisotropic cosmological metric ansatz Eq.~\eqref{KS} of Kantowski-Sachs becomes increasingly worse. In other words, such an ansatz breaks down while approaching the singularity and the spacetime is fractured into small parts according to the BKL conjecture \cite{Belinsky:1982pk}. Rather than having a smooth approach towards the singularity, BKL stipulates that the collapse is generically chaotic and leads to the individual sub-regions not interacting with each other.

This fact has been recently incorporated by Perry \cite{Perry:2021mch} to study the feasibility of imposing the DeWitt boundary condition for a vacuum black hole spacetime (see also \cite{Kleinschmidt:2009cv}). Here, we briefly restate his arguments for completeness. According to \cite{Perry:2021mch}, the wave function for each of these small parts, following a Kasner spacetime metric, will satisfy the Wheeler-DeWitt equation corresponding to a Hamiltonian constraint given by:
\begin{eqnarray}
H = \frac{1}{2} \left( \pi_{\rho}^{2} + \frac{v^{2}}{\rho^{2}} \left( \pi_{u}^{2} + \pi_{v}^{2} \right) \right)\,,
\end{eqnarray}
where $\pi_{\rho,u,v}$ are canonical momenta of the coordinates $\rho$, $u$, $v$, respectively. Furthermore, dynamics is restricted within a spatial sub-volume such that one must impose vanishing boundary conditions for this compact spatial region over the $uv$-hypersurface. Due to this, the eigenvalues for the spatial direction $\Delta$ is positive definite and
\begin{eqnarray}
\left( - \rho^{2} \frac{d^{2}}{d\rho^{2}} - 2\rho \frac{d}{d\rho} - \Delta \right) \Psi = 0
\end{eqnarray}
along the direction $\rho$, where the increasing $\rho$ direction corresponds to the increasing time direction. As we solve the generic solution of $\Psi$, the wave function must go as some inverse-power of $\rho$ which goes to infinity at the singularity. Therefore, the wave function for each part necessarily goes to zero and resembles precisely the condition due to DeWitt. Thus, considering only the Wheeler-DeWitt equation, one finds that the wave function can go to zero at the singularity provided we are willing to change the metric ansatz of the spacetime within the black hole, as one approaches the singularity, according to the BKL conjecture. 

In conclusion, if the BKL conjecture is true and one imposes the vanishing boundary condition for the local spatial boundary, the wave function should vanish at the singularity. Note that one does not actually impose the DeWitt boundary condition in this case as an additional input; rather, the BKL conjecture naturally leads to the DeWitt condition on suitably modifying the Wheeler-DeWitt equation as one approaches the singularity.

Given this nice result, one can ask two questions:
\begin{itemize}
\item[1.] Can we generalize this result to find the DeWitt boundary condition beyond the vacuum case for the more realistic scenario of gravitational collapse?
\item[2.] Is the DeWitt boundary condition consistent with other quantum gravity approaches, \textit{e.g.}, loop quantum gravity?
\end{itemize}
As we shall demonstrate in the rest of the paper, we will indeed find the answer to both of these questions to be in the affirmative.

\section{\label{sec:dewitt}DeWitt boundary condition for collapsing matter}
\subsection{Hajicek-Kiefer wave packet}
Following Hajicek and Kiefer \cite{Hajicek:2000mh}, we start from the action for a null shell
\begin{eqnarray}
S = \int d\tau \left( p_{u} \dot{u} + p_{v} \dot{v} - n p_{u} p_{v} \right)\,,
\end{eqnarray}
where $u$ is the in-going null direction, $v$ is the out-going null direction, $p_{u}$ and $p_{v}$ are conjugate momenta of $u$ and $v$, respectively, and $n$ is a Lagrange multiplier, where $p_{u}p_{v}=0$ is the necessary constraint. On quantizing the shell, we introduce the inner product
\begin{eqnarray}
\langle \phi | \psi \rangle = \int_{0}^{\infty} \frac{dp}{p}\, \phi^{*}(p)\, \psi(p)\, ,
\end{eqnarray}
where this $p$ is a continuous variable for the representation. Thanks to this choice of the inner product, after some computations, one can show that the eigenvalue of the operator $p_{t} = p_{u} + p_{v}$ is $-p$. Hence, this $p$, the eigenvalue that conjugates to the time variable, can be interpreted by the total energy of the system (see more details in \cite{Hajicek:2000mi}).

Given the radius $r = (-u + v)/2$, we construct the radial operator
\begin{eqnarray}
\hat{r}^{2} = - \sqrt{p}\, \frac{d^{2}}{dp^{2}}\, \frac{1}{\sqrt{p}}\,.
\end{eqnarray}
Then, we obtain the relation $\hat{r}^{2} \langle r | p \rangle = r^{2} \langle r | p \rangle$. From this, we find:
\begin{eqnarray}
\langle r | p \rangle = \sqrt{\frac{2p}{\pi}}\, \sin rp\,.
\end{eqnarray}

Therefore, for a given wave packet $\phi(p)$, the normalization condition translates into
\begin{eqnarray}
1 = \int_{0}^{\infty} \frac{dp}{p} \left| \phi(p) \right|^{2}\,.
\end{eqnarray}
Note that $t = (u+v)/2$. After the integral transformation of the time-dependent wave function $\phi(p) e^{-ipt}$, we obtain
\begin{eqnarray}
\Psi(t,r) = \sqrt{\frac{2}{\pi}} \int_{0}^{\infty} \frac{dp}{\sqrt{p}}\, \phi(p)\, e^{-ipt}\, \sin rp\,,
\end{eqnarray}
which is again a normalized wave function.

\begin{figure}[h]
\begin{center}
\includegraphics[scale=0.7]{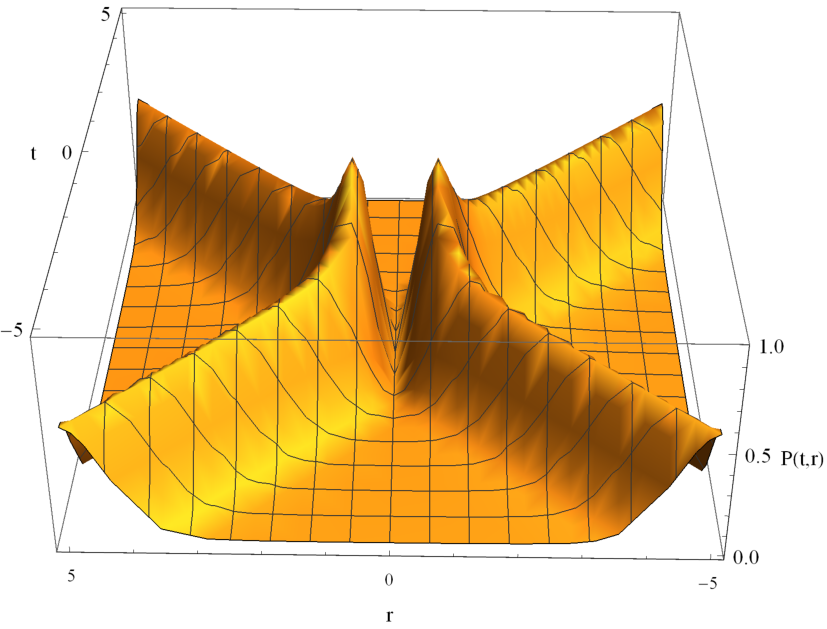}
\caption{\label{fig:Gaussian}$|\Psi|^{2}$ of the Gaussian wave packet form (Eq.~(\ref{eq:G})) for $p_{0} = \sigma = 1$.}
\includegraphics[scale=0.7]{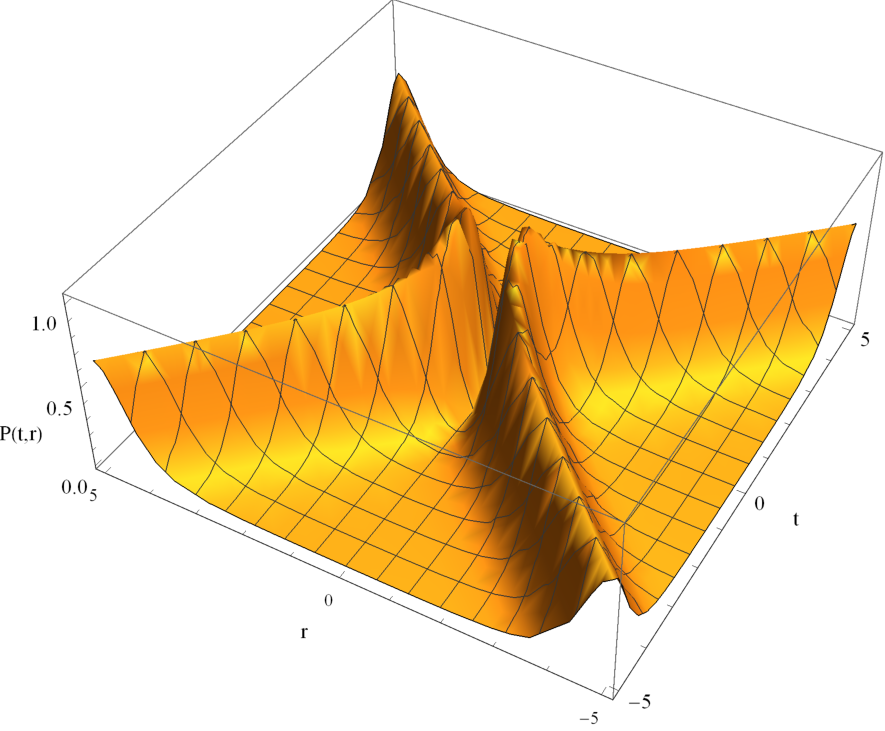}
\caption{\label{fig:plot}$|\Psi|^{2}$ of original Hajicek-Kiefer paper (Eq.~(\ref{HKwave})) for $\lambda = \kappa = 1$.}
\end{center}
\end{figure}

We can take two explicit examples:
\begin{itemize}
\item[1.] If we choose the normalized Gaussian wave packet (Fig.~\ref{fig:Gaussian})
\begin{eqnarray}\label{eq:G}
\phi(p) = \left(\frac{2}{\pi \sigma^{2}}\right)^{1/4} \left( 1 + \mathrm{erf}\left( \frac{p_{0}}{\sqrt{2}\sigma}\right) \right)^{-1/2} \sqrt{p}\, e^{- \frac{\left( p - p_{0} \right)^{2}}{4\sigma^{2}}}\,,
\end{eqnarray}
after the integral transformation, we obtain a bouncing condition at the singularity, where $p_{0}$ is the peak of the Gaussian wave function, $\sigma$ is the standard deviation in the $p$-space, and $\mathrm{erf}$ is the error function.
\item[2.] In the original Hajicek-Kiefer paper, the following wave packet was introduced (Fig.~\ref{fig:plot}):
\begin{eqnarray}\label{wavepacket}
\phi(p) = \frac{(2\lambda)^{\kappa + 1/2}}{\sqrt{(2\kappa)!}}\, p^{\kappa + 1/2}\, e^{-\lambda p}\,.
\end{eqnarray}
For this case, we obtain the following analytic form of the wave function
\begin{eqnarray}
\Psi_{\kappa \lambda} \left( t, r \right) = \frac{1}{\sqrt{2\pi}} \frac{\kappa! (2\lambda)^{\kappa+1/2}}{\sqrt{(2\kappa)!}} \left[ \frac{i}{\left( \lambda + it + ir \right)^{\kappa + 1}} - \frac{i}{\left( \lambda + it - ir \right)^{\kappa + 1}} \right],\label{HKwave}
\end{eqnarray}
where $\kappa$ is a positive integer and $\lambda$ is a positive number.
\end{itemize}

For both cases, the broad physical properties are the same. If we evaluate the probability $|\Psi|^{2}$, which is shown respectively in Fig.~\ref{fig:Gaussian} and Fig.~\ref{fig:plot} for the above two examples, the steepest-descent has four classical branches: (1) $t = - r$ for $t \rightarrow -\infty$, (2) $t = r$ for $t\rightarrow +\infty$, (3) $t = -r$ for $t \rightarrow +\infty$, and (4) $t = r$ for $t \rightarrow - \infty$. Here, (1) and (2) are connected via a quantum bounce region, and (3) and (4) are connected via another quantum bounce region. (1) and (2) are disconnected to (3) and (4) because of the destructive superposition of waves. In addition, note that if the mass of the shell is greater than the Planck scale, the classical part of the shell can cross the event horizon.

\subsection{Interpretation: shell-antishell pair annihilation}
The first observation is that what we considered above is not the quantization of spacetime, but rather the quantization of the shell. This is the reason why $t$ and $r$ are coordinate variables. In particular, there is another notion of time, called causal time $\tau$. In the standard Penrose diagram, the forward direction of causal time $\tau$ always points upward. In fact, the coordinate variables $t$ and $r$ can denote the field value if they denote the location of the shell. Therefore, this model cannot certainly resolve the space-like singularity.

\begin{figure}
\begin{center}
\includegraphics[scale=1.6]{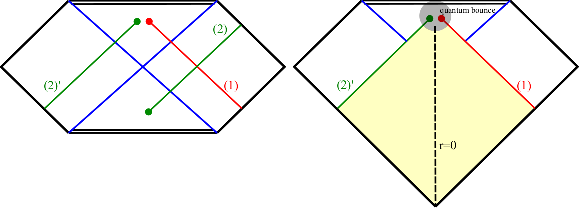}
\caption{\label{fig:HK}Left: possible trajectories for (1) $t = - r$ for $t \rightarrow -\infty$ and (2) $t = r$ for $t\rightarrow +\infty$ on the Penrose diagram. The second branch can be interpreted in two ways: shell moving forward in causal time $\tau$ or moving backward in $\tau$. Right: the only physically sensible connection of two branches is to connect (1) and (2)$'$. The yellow-colored region corresponds to the Minkowski spacetime.}
\end{center}
\end{figure}

Therefore, in Fig.~\ref{fig:HK}, it is reasonable to interpret the trajectory (1) as the collapsing shell. However, due to the ambiguity of the arrow of time, the trajectory (2) can be interpreted in two ways. One is that a positive tension shell moves from the white hole region to the infinity forward both in its causal and coordinate time (denoted by (2) in Fig.~\ref{fig:HK}); the other is a negative tension shell moves from a black hole to the past infinity backward in its causal time $\tau$, but forward in coordinate time $t$ (denoted by (2)$'$ in Fig.~\ref{fig:HK}).

Then the solution shows that two classical branches could collide as a quantum bounce. Before and after the quantum bounce, which branches are connected? One possibility is to connect from (1) to (2) in the Left of Fig.~\ref{fig:HK}. The issue is that this would require the shell to cross the space-like singularity. However, as we mentioned, this approach does not quantize the spacetime itself; hence, it cannot change the structure of the space-like singularity.

The only sensible interpretation is to interpret that the second branch (2)$'$ moves backward in its coordinate time. This is not impossible because there is no intrinsic arrow of time. For (1), we assume that the initial condition corresponds to the collapsing shell from the past infinity; hence, we can choose this by construction. However, for (2), there is no such a principle and it is still consistent to connect (1) to (2)$'$.

This implies that as the positive tension shell collapses, there exists a \textit{counter-shell} which has the negative tension but locates beyond the Einstein-Rosen bridge. Two shells then collide inside the event horizon; this corresponds to the quantum bounce. Two shells are \textit{annihilated} and eventually a pure Schwarzschild black hole remains. (For the shell-anti-shell pair creation of the time-like shell, see \cite{Chen:2020nag}.)

\begin{figure}[h]
\begin{center}
\includegraphics[scale=0.5]{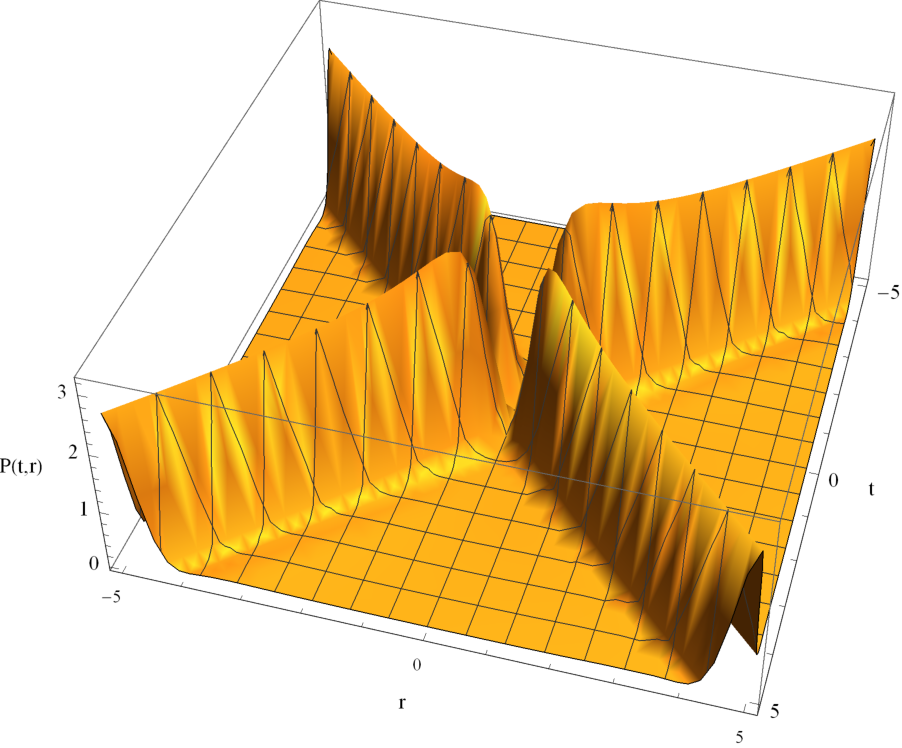}
\includegraphics[scale=1.6]{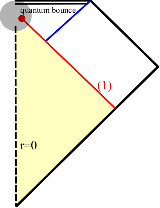}
\caption{\label{fig:HK3}Left: $|\Psi'|^{2}$ for $\lambda = \kappa = 1$. Right: The physical interpretation of $\Psi'$.}\label{wave3}
\end{center}
\end{figure}

The existence of the anti-shell is not necessarily explicit. Since $\Psi_{\kappa \lambda} \left( t, r \right)$ is the wave packet, one can further consider the following superposition, in the meantime maintaining the destructive behavior near the singularity:
\begin{eqnarray}
\Psi'_{\kappa \lambda} \left( t, r \right) \propto \Psi_{\kappa \lambda} \left( t, r \right) - \Psi_{\kappa \lambda} \left( -t, r \right).
\end{eqnarray}
The probability for this superposition is shown in Fig.~\ref{wave3}. Now four classical branches are all disconnected. There exists an anti-shell contribution that provides a destructive superposition, but it is hidden beyond the $r=0$ boundary.

\subsection{Extending this interpretation to the black hole spacetime itself?}

The Hajicek-Kiefer wave function describes the vanishing boundary condition near the singularity; this can be interpreted as a destructive interference between the matter shell and the anti-matter shell. The quantized wave function is symmetric up to the change of $t \rightarrow - t$ and $r \rightarrow - r$, and hence, if there is a steepest-descent of a collapsing matter shell, there must exist a steepest-descent of an anti-matter shell. The collision of the two shells makes the vanishing boundary condition at $r = 0$ inevitable.

Therefore, the origin of the vanishing boundary condition of the matter shell is due to the reflection symmetry in space and time itself. Indeed, this is not surprising because if we impose the vanishing boundary condition, it is equivalent to say that there exists a symmetric pulse that induces the destructive interference. If this correspondence is true for the collapsing shell, then perhaps it is also true for the black hole geometry itself. In other words, can we interpret the DeWitt boundary condition of the black hole as a destructive interference due to the symmetric anti-geometry? Indeed, as we shall now show, this can be explicitly realized within models of loop quantum gravity.

\section{\label{sec:black} Black holes and the DeWitt condition in loop quantum gravity}
Loop quantum gravity is a non-perturbative quantum gravitational theory based on the canonical quantization approach. It is characterized by the discretized spectra of the geometrical operators that constitute the spacetime from the quantum point of view. A direct consequence of such discretized spectra is that there exists a fundamental area gap in the theory. In the semi-classical regime, effective loop quantum gravity models can be constructed in ways that the quantum area gap plays a crucial role in introducing Planckian nonlocalities such that curvatures become bounded, and therefore, sheds light on ameliorating classical singularities in generic black hole spacetimes.

\subsection{Brief history of understanding black holes in loop quantum gravity}
In this subsection, we give a very brief (historical) review of the status of black holes in the loop quantum gravity community. There have been several paradigm changes, but there is a constant common factor: {\it quantum-geometry effects} typically result in new \textit{spacetime symmetries} which eventually guide the causal structure. Let us quickly explain what we mean by this statement. As we shall show, there are two types of models seemingly coming out of loop quantum gravity. The first class of these models posit a black hole bounce into a white hole, where there a \textit{symmetry} exists between the black hole and the white hole geometries which will play a crucial role in our analysis going forward. Another class of models not only consider the homogeneous interior, but also take into account the modified gauge transformations which result from the holonomy corrections implied by loop quantum gravity. In these models, one finds a Euclidean core in the interior of the black hole due to \textit{dynamical signature-change}, resulting in a new spacetime structure, which follows from the deformed symmetries of the theory. In summary, quantum gravity effects are crucial for determining the spacetime symmetries and causal structure of black holes in loop quantum gravity.

\begin{figure}
\begin{center}
\includegraphics[scale=0.3]{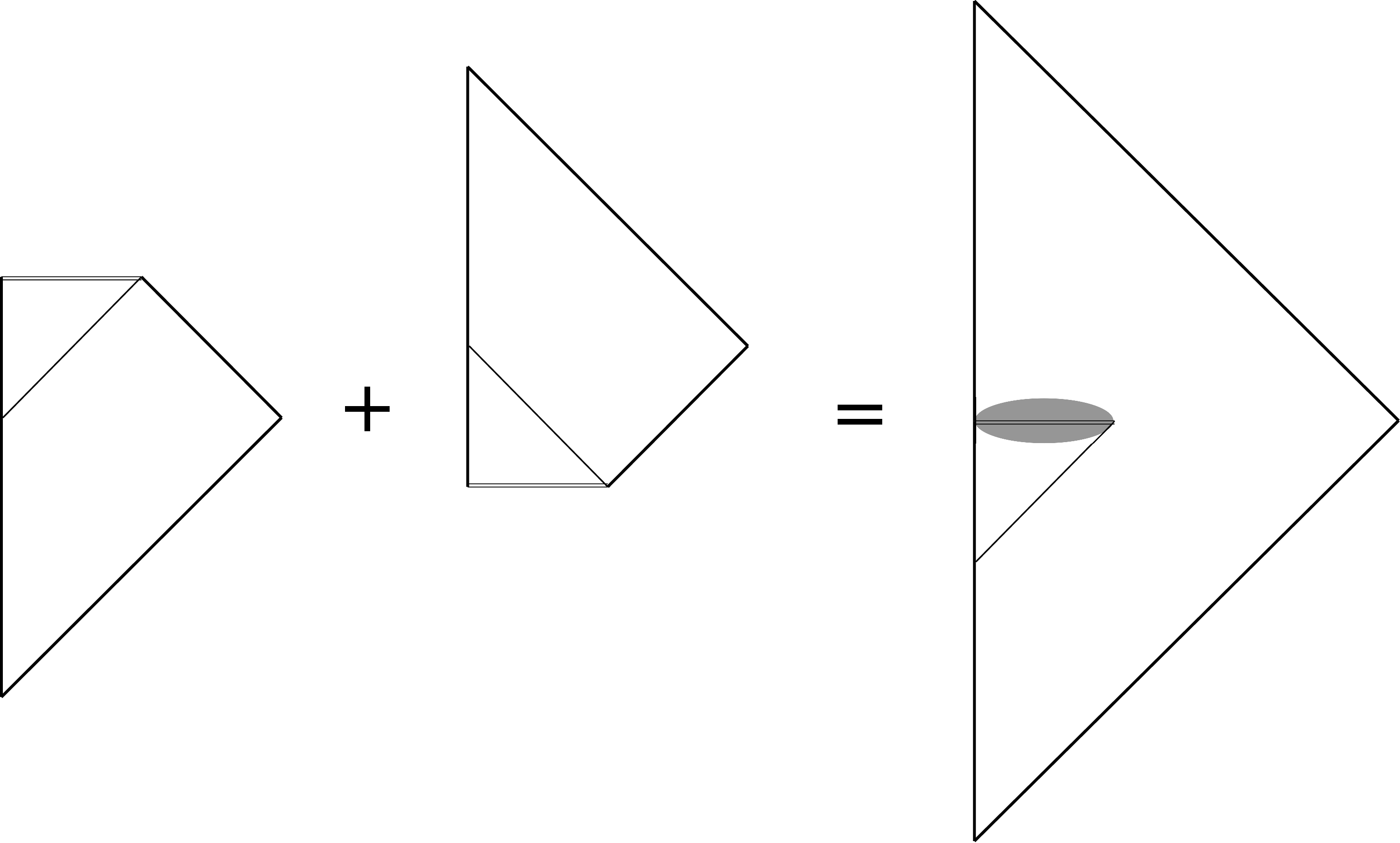}
\caption{\label{fig:AB} Ashtekar-Bojowald model: The gray colored region is the place where quantum gravitational treatments applied.}
\end{center}
\end{figure}

\subsubsection{Ashtekar-Bojowald model}
In the loop quantum gravity community, the first major effort to understand the evolution of black holes, from gravitational collapse to complete evaporation, was introduced by Ashtekar and Bojowald \cite{Ashtekar:2005cj} (Fig.~\ref{fig:AB}). If the space-like singularity inside a black hole can indeed be resolved by quantum gravitational effects, then the global causal structure must be connected in some way. Ashtekar and Bojowald emphasized the symmetry between the black hole phase and the white hole phase. The main input is that although these two black hole and white hole phases are classically disconnected, quantum geometry corrections can lead to them being smoothly connected akin to what happens in bouncing models of loop quantum cosmology. Although the concrete model in which the singularity-resolution was explicitly demonstrated was purely geometric without the addition of any matter (and due to spherical symmetry, lacked dynamical graviton degrees of freedom as well), it was argued that the main qualitative features of singularity-resolution would also hold when considering matter fields and removing the restriction of being confined to the interior of the Schwarzschild spacetime. In order to draw the complete global causal structure, the main task is how to connect the infinity. In their picture, they connected past and future infinities with the physical motivation that the black hole will evaporate in a finite amount of time. Therefore, in terms of the information loss paradox, it was argued that since the black hole evaporates in a finite amount of time, the future null infinity {\it lies} to the future of the deep quantum regime where the classical singularity originally was, and there are family of observers who shall recover the apparently `lost' information. These observers reside entirely in the asymptotic region going from the past null infinity to the future null inifinity and they never go near the deep quantum regime. This picture is somewhat similar to what has been discussed by Stephens, 't~Hooft and Whiting \cite{Stephens:1993an}. 


Nowadays, however, we think that this picture is incomplete and it does not resolve the information loss paradox. Firstly, if almost all of the information is squeezed near the singularity and Hawking radiation carries almost no information, then the entropy of the quantum gravitational region must be almost infinite, and this, in turn, suffers from the typical problems of black hole remnants \cite{Chen:2014jwq}. On the other hand, even if Hawking radiation carries information away in some way, it cannot still rescue us from the problem of information duplication \cite{Yeom:2008qw}. There have been several trials to demonstrate the Ashtekar-Bojowald-like spacetime based on regular black holes \cite{Ayon-Beato:1999kuh}; however, it is very subtle whether their metric ansatz can be justified even for dynamical cases beyond the Vaidya-like approximation \cite{Brahma:2019oal}. Finally, the holonomy corrections which lead to singularity-resolution in loop quantum gravity in the first place, seem to severely deform the spacetime geometry so that the black hole has an Euclidean core (known as \textit{dynamical signature-change} \cite{Sig_Change}) where Cauchy evolution breaks down \cite{Bojowald_criticism}. We shall briefly come back to this point later on.

Despite these criticisms, it is worth mentioning that the Ashtekar-Bojowald picture emphasizes two issues: (1) we need to extend beyond the singularity and ask how can we connect the interior to the outside spacetime and (2) the spacetime symmetry can guide us towards a solution for the problem. Notwithstanding the issue of resolving the information loss paradox, and the problem of signature-change, the Ashtekar-Bojowald picture does provide a fresh perspective on singularity-resolution in black hole spacetimes.

\begin{figure}
\begin{center}
\includegraphics[scale=0.3]{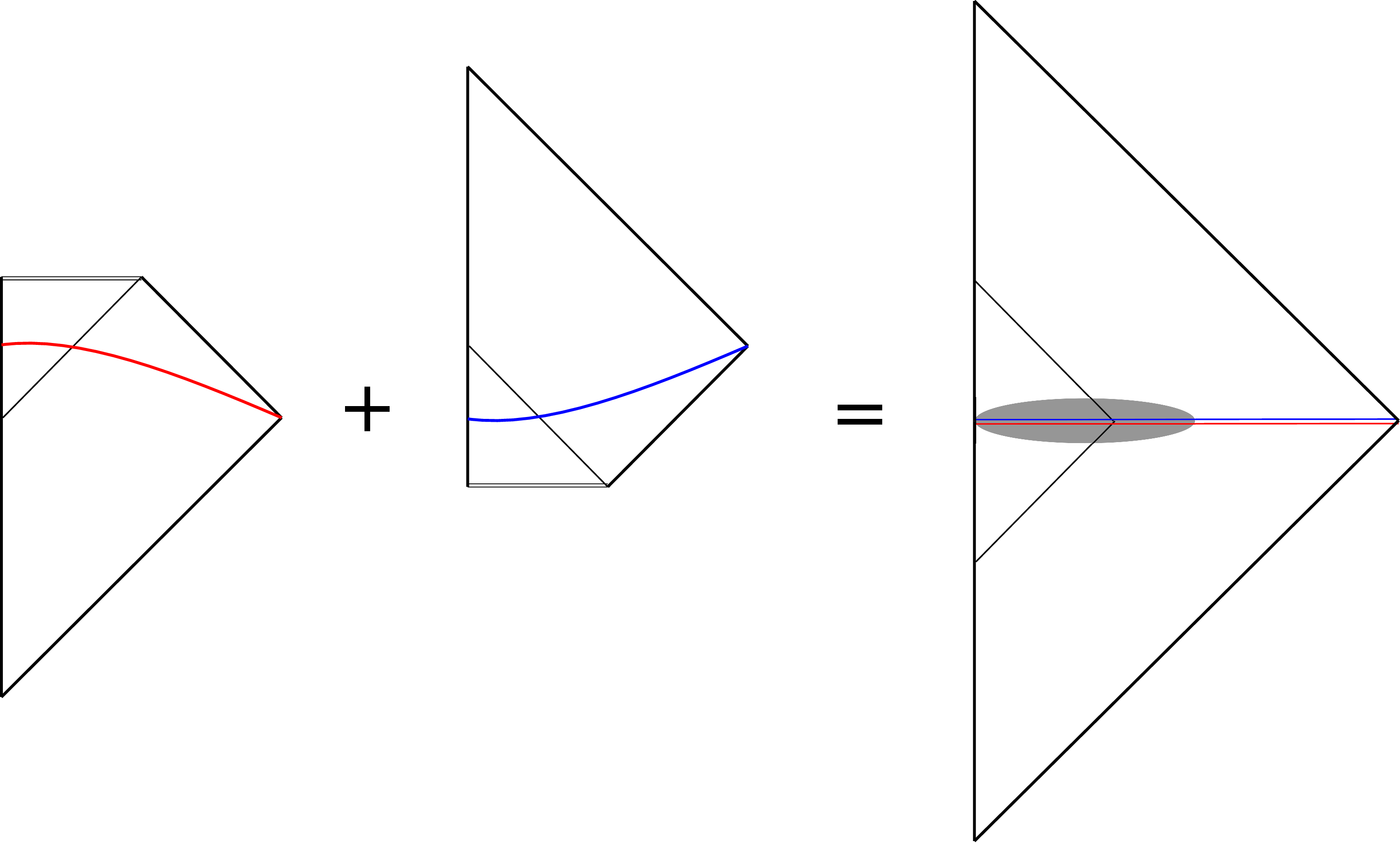}
\caption{\label{fig:HR}Haggard-Rovelli model. We cut the lower part of the red curve (left) and the upper part of the blue curve (middle); finally, paste them (right). The gray colored region is the place where quantum gravitational treatments applied; this reaches outside the event horizon.}
\end{center}
\end{figure}

\subsubsection{Black hole fireworks}
In order to connect inside to the exterior of the black hole, Haggard and Rovelli proposed a more radical idea \cite{Haggard:2014rza} (Fig.~\ref{fig:HR}). After resolving the singularity, we need to connect from inside to outside; in other words, to connect from a black hole phase to a white hole phase. However, in order to cut-and-paste the spacetime smoothly, Haggard and Rovelli cut and paste not only inside the event horizon, but also outside the event horizon. This can, of course, only work if we believe that quantum gravity effects can leak outside the event horizon. The authors named this process as a \textit{black hole firework} or a \textit{horizon spark}.

There are several problems in this picture. First, despite some efforts from the spin-foam approach, there is still no good mechanism to justify the firework from first principles. Like the firewall picture \cite{Almheiri:2012rt}, general relativity must be violated even near the horizon scale, where these effects can be quite pronounced \cite{Kim:2013fv}. Also, as we rely on the cut-and-paste technique, it is inevitable that this leads to finding quantum corrections not only near the horizon, but also asymptotically away from it near the infinity \cite{Brahma:2018cgr}. And this conclusion not only stands for this particular model of \cite{Haggard:2014rza}, but more generally to such black-to-white hole transitions which can originate from shock waves or other similar effects \cite{Shock_Waves}. 

One may think this model may be too adventurous an idea, but in any case, this remains a logical possibility to connect from a black hole phase to a white hole phase. However, note here that connecting two such spacetimes in the presence of any realistic matter component, while assuming continuity of the spatial metric (one of the Israel junction conditions), typically indicates that either there must be some quantum gravity effects leaking outside the horizon or one must have an inner horizon developed in the dynamical process of such a transition \cite{BtoW_NG}.

\begin{figure}
\begin{center}
\includegraphics[scale=0.3]{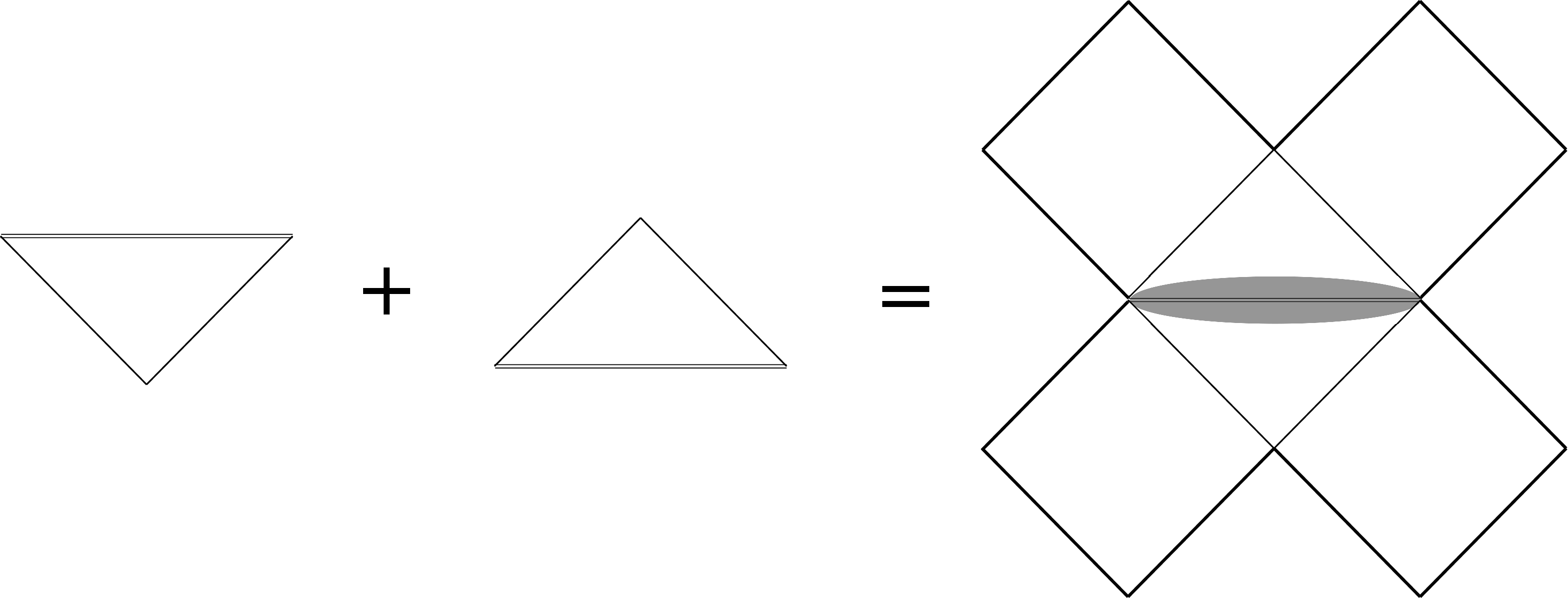}
\caption{\label{fig:AOS}Ashtekar-Olmedo-Singh/Bodendorfer-Mele-Munch model.}
\end{center}
\end{figure}

\subsubsection{Homogeneous bouncing black hole models}\label{Bounce}
Another possibility which remains is to extend the spacetime inside the horizon, while we do not explicitly connect the internal spacetime with the future infinity. In this case, the black hole geometry is explicitly connected to the white hole spacetime only inside the horizon. The advantage of this picture is that it is mathematically very clean and such a bouncing geometry can be explicitly derived within minisuperspace models of loop quantum gravity. (Some of the problems of only considering the interior of the Schwarzschild spacetime as an anisotropic cosmological model and applying techniques of loop quantum cosmology have been illustrated in \cite{BtoW_NG, Bojowald_criticism, Bojowald_NonCov}.) Nevertheless, depending on quantization ambiguities, one can get such a bounce with two different possibilities, \textit{i.e.}, with \cite{Bojowald:2018xxu, BenAchour:2018khr} or without the existence of an inner apparent horizon \cite{Ashtekar:2018lag, Bodendorfer:2019xbp}.

These types of bounces are essentially obtained when treating the interior as a homogeneous spacetime and other complications arise when considering deformations of the Dirac algebra coming from quantum geometry corrections. Keeping aside this subtlety for now, note that even for the black-to-white hole bounce considered within the anisotropic cosmology picture, different models relying on `polymerizing' different variables would lead to a distinct outcomes. For instance, in \cite{Ashtekar:2018lag} a model was proposed which has only one Dirac observable (the black hole mass being essentially the same as the white hole one) (Fig.~\ref{fig:AOS}), whose consistency has subsequently been questioned from different perspectives \cite{Bouhmadi-Lopez:2019hpp}, but one can consider another model with a similar causal structure \cite{Bodendorfer:2019xbp} (and its refinements \cite{Bodendorfer2}) which seem to be much more consistent \cite{Bouhmadi-Lopez:2020oia}.



Even if we first accept this approach, and come back to its drawbacks later on, there are still several unanswered questions. What is the causal structure for the evaporating spacetime? What is the destiny of the information inside the black hole? Is the white hole horizon stable against perturbations? Nevertheless, despite multiple shortcomings, this remains one of the most popular descriptions of curing the singularity, and extend the black hole spacetime, within loop quantum gravity \cite{Gambini:2013ooa}. 

\begin{figure}
\begin{center}
\includegraphics[scale=0.3]{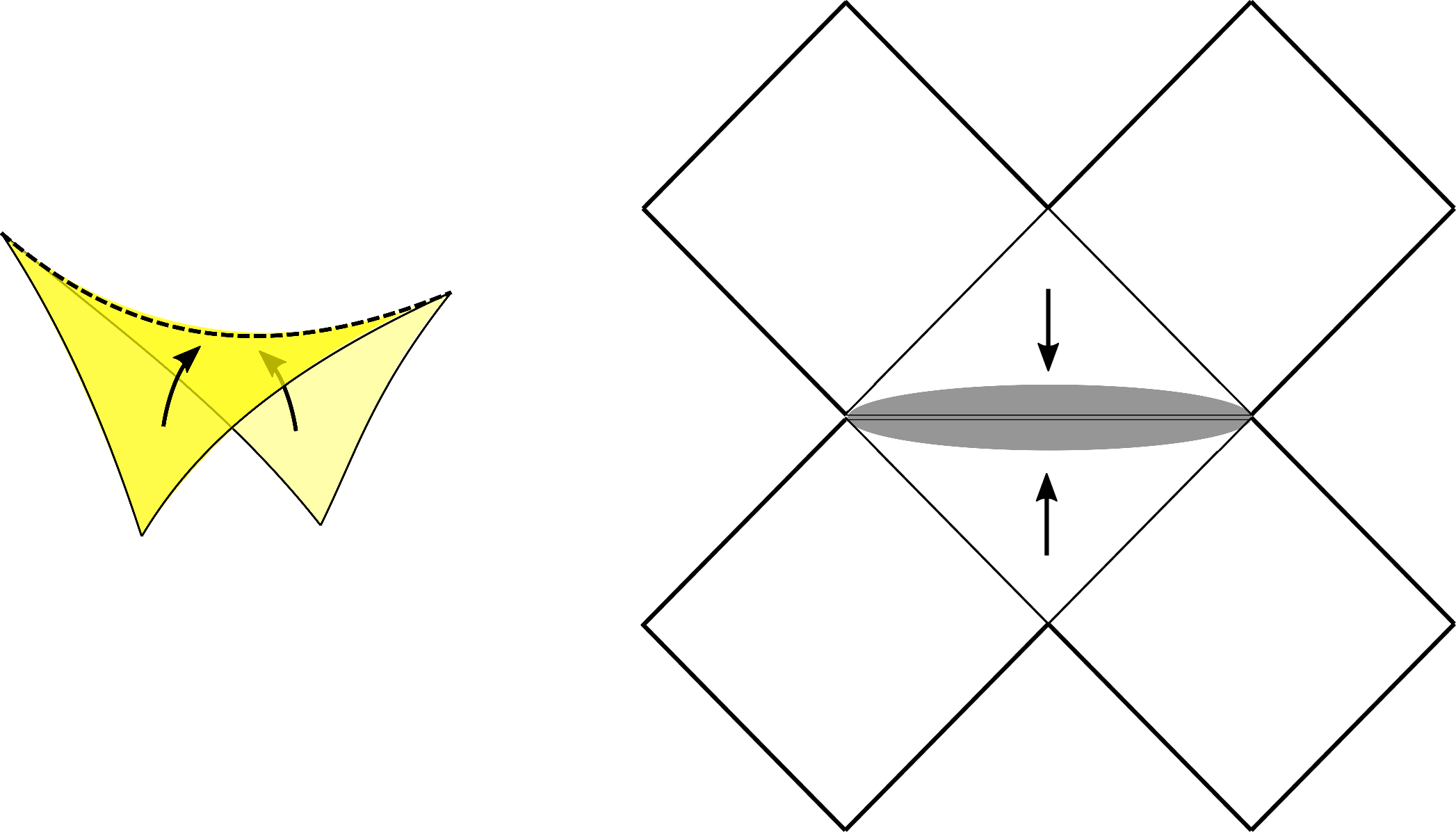}
\caption{\label{fig:AN}Annihilation-to-nothing interpretation. Left: two black hole phases collide and are annihilated. Right: physical arrows of time can be flipped around the quantum gravitational region (gray colored region).}
\end{center}
\end{figure}

\subsubsection{Introducing a new paradigm: annihilation-to-nothing}
Above, we have briefly reviewed some of the important approaches often used to explore the black hole spacetime in loop quantum gravity. The main message is that quantum gravity effects, mainly in the form of holonomy corrections which replace unbounded curvature components by bounded functions of them, will resolve the classical singularity. Even if singularity in the form of diverging curvatures is avoided, the important question that remains is the fate of the causal structure of the black hole spacetime due to such quantum geometry effects. How can the spacetime be extended `beyond' the singularity? A popular answer is that the geometry beyond the singularity is something like the time-reversal of the black hole geometry; in other words, something which resembles a white hole. However, how can we connect the past infinity and the future infinity? If two infinities meet each other (Ashtekar-Bojowald paradigm or the fireworks picture), they suffer from problems regarding the information loss paradox. If two infinities do not meet each other (as in the bouncing models in Sec.~\ref{Bounce} above), not only may one already lose the information, but also the future of the collapsing matter is much less clear.

However, up until now, for all these paradigms, one necessarily accepts that \textit{there exists only one arrow of time}. However, as soon as one admits that there exists a quantum gravitational region where a semi-classical description of geometry is unavailable, the notion of the arrow of time becomes ambiguous and unclear. In fact, one can introduce \textit{two arrows of time} before and after the quantum bouncing point (in the cosmological context, see \cite{Chen:2016ask}). This may open a new way to understand the spacetime diagram of such quantum black holes (Right of Fig.~\ref{fig:AN}).

Then in this picture, indeed, there are not one black hole and one white hole geometries, but rather two black hole phases. The two black hole phases collide at the putative singularity (Left of Fig.~\ref{fig:AN}). In other words, we have destructive interference between these two geomtries such that there is no outcome of the collision and the two black hole phases get completely annihilated at this surface. This is the inspiration behind naming this paradigm the \textit{annihilation-to-nothing} interpretation \cite{Bouhmadi-Lopez:2019kkt}. After the collision, the wave function must completely vanish and therefore, at this point, this interpretation is naturally connected to the DeWitt boundary condition. In this case, the crucial new ingredient for us is the ability to utilize the ambiguity in defining the arrow of time, so as to have two arrows of time in the quantum gravity setup, such that the wave function vanishes at the ``bouncing'' hypersurface. Let us reiterate that we do not directly invoke the DeWitt boundary condition but rather find it as a consequence of introducing the new arrow of time. Finally, note that typically in loop quantum gravity, the {\it transition hypersurface} is space-like and this is what replaces the classical singularity, and for our new interpretation, it is on this transition hypersurface that the wave function vanishes and the DeWitt condition is naturally recovered.

The reader might be confused at this point as to how it is now that we find the vanishing of the wave function on this transition hypersurface, even after imposing classicality at the horizon. As mentioned earlier, in the case of the usual Wheeler-DeWitt equation, one has to modify the Kantowski-Sachs metric to a different Kasner spacetime in order to do so. For the case of loop quantum black holes, this is a consequence of having different evolution equations -- the so-called effective equations -- which have large deviations from the standard Wheeler-DeWitt equation near Planck scales. This is why something strongly reminiscent of the DeWitt boundary condition can be found in loop quantum gravity, given our new \textit{annihilation-to-nothing} interpretation.

\subsection{Embedding collapsing shells into loop quantum black holes}
As mentioned above, in the case of black holes, the space-like singularity in the classical Schwarzschild black hole is shown to be avoidable in several effective loop quantum black hole models. Although the method of constructing effective black hole models is not unique and there are several possibilities depending on quantization ambiguities, one common feature in the bouncing models is that the classical singularity can be replaced by some sorts of transition surface, or a bouncing surface inside the horizon. This kind of effective spacetime structure may be connected to the picture of the vanishing wave function mentioned in the previous section.

Generically, the spacetime metric of a loop quantum black hole that has an interior bouncing surface can be described using the following metric (we choose the explicit expression in \cite{Bodendorfer:2019xbp} for concreteness):
\begin{equation}
ds^2=-\left[1-\frac{2M(y)}{b(y)}\right]dt^2+\left[1-\frac{2M(y)}{b(y)}\right]^{-1}dy^2+b(y)^2d\Omega^2\,,
\end{equation}
where $b(y)$ is the areal radius and it is a function of another radial variable $y$. The mass function $M(y)$ is not a constant anymore, but it should reduce to the ADM mass in the asymptotic region. The bouncing feature can be seen on realizing that the areal radius $b$ has a non-zero minimum value at $y=0$. In the usual interpretation, the black hole region corresponds to a positive $y$. After crossing the transition surface, the spacetime has a negative $y$ and the spacetime bounces to the white hole region. 

\begin{figure}
\begin{center}
\includegraphics[scale=1.6]{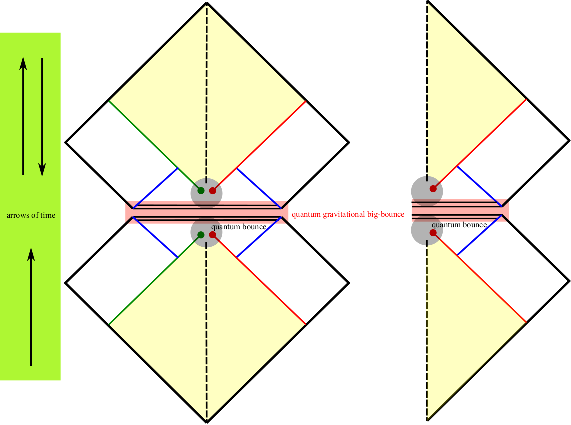}
\caption{\label{fig:HK4}If there is any principle to explain a quantum big-bounce near the singularity, one can accordingly explain the destination of the collapsing shell.}
\label{FIG.HK4}
\end{center}
\end{figure}

If one applies the quantized null shell picture of Hajicek-Kiefer to the bouncing loop quantum black hole model, one may naively write down $y=(-u+v)/2$ and the radial operator is associated with $\hat{y}$:
\begin{equation}
\hat{y}^2=-\sqrt{p}\frac{d^2}{dp^2}\frac{1}{\sqrt{p}}.
\end{equation}
The eigenfunction $\langle y|p\rangle$ becomes
\begin{equation}
\langle y|p\rangle=\sqrt{\frac{2p}{\pi}}\sin{yp}.
\end{equation}
In this regard, the time-dependent wave function, after imposing the wave packet Eq.~\eqref{wavepacket}, can be expressed as that given in Eq.~\eqref{HKwave}, while replacing all $r$ with $y$. Therefore, the wave function vanishes at the bouncing point $y=0$. The corresponding spacetime diagrams are exhibited in Fig.~\ref{FIG.HK4}.

Here, let us strongly emphasize that this is a very heuristic picture of including matter shell collapse into black hole models of loop quantum gravity. In fact, one of the main outstanding tasks is to find a closed algebra of constraints when adding matter to spherically symmetric models of loop quantum gravity. This is, of course, related to treating the entire black hole spacetime and not just focusing on the interior of the horizon, to which class of models the homogeneous bounce models mentioned above reside. Thus, in order to derive a full treatment of matter shells collapse in loop quantum gravity, much is yet to be done. (For some preliminary efforts in this direction, see \cite{Kelly:2020lec}.) But our above qualitative argument gives us some hope that once this task is carried out, it will still be possible to extend the DeWitt boundary condition for such matter shell collapse models in loop quantum gravity as well.

\subsection{New interpretation in the light of the DeWitt boundary condition}

The putative singularity disappears due to the holonomy corrections in loop quantum gravity. However, in the deep quantum regime when these quantum geometry effects take over, it is not strange to think that we loose a definite direction of time. Here, interestingly, we can find several pieces of quantum-modified spacetime, where one can impose different arrows of time for each piece. So, if we believe that there exists only one arrow of time, the gravitational collapse and the geometry inside the horizon smoothly continue onto the bouncing matter as well as the bouncing white hole geometry. On the other hand, if we assume that there are two arrows of time, then the two gravitational collapsing black hole geometries are annihilated to nothing; at this annihilation surface, there is a quantum bounce (in the point of view of loop quantum gravity) or the vanishing boundary condition (\textit{i.e.}, the DeWitt boundary condition). In other words, as two geometries approach together, they are annihilated, and become really quantum. In conclusion, the DeWitt boundary condition can be consistent with the loop quantum gravity inspired black hole models, although there is an extra ingredient in the form of interpretation of the arrows of time.

Let us elaborate a bit now on a common problem associated with these paradigms in loop quantum gravity. Typically, one assumes a pivotal role of these so-called holonomy corrections (which result from choosing a specific regularization scheme of the Hamiltonian constraint operator) in the effective equations of motion governing the dynamics of the black hole. On the other hand, the effects of such quantum geometry corrections on the spacetime symmetries have remain relatively less-explored. Recently, it has been shown that the same holonomy modifications which result in singularity-resolution in loop quantum gravity, also lead to deformation of the diffeomorphism symmetry of the background, as evidenced from the Dirac constraint algebra \cite{Sig_Change}. In other words, Riemannian line-elements are not sufficient to describe the quantum geometry near the deep quantum regimes in loop quantum gravity and one has to reconcile the holonomy-corrected Hamiltonian constraint operator with a deformed notion of covariance \cite{Bojowald:2016hgh}.  One has to suitably modify the classical line-elements and the classical gauge conditions, as would be appropriate for gauge-transformations generated by the quantum-corrected Hamiltonian constraint operator \cite{Bojowald:2018xxu}. The bottom line is that incorporating the effects of such modified gauge-transformations typically results in a dynamical signature-change of the metric and one ends up with a Euclidean core in the deep quantum regime of the loop-quantized black hole \cite{Bojowald_criticism}.

In light of the DeWitt boundary condition, this, in fact, turns one of the weaknesses of bouncing black hole models of loop quantum gravity into a strength. To see how this comes about, first note that in the context of the gravitational path integral, it is not at all straightforward to distinguish between the past and the future time slices. This, of course, is related to the ambiguity of the arrow of time mentioned earlier. On the other hand, if we take the approach of the Hartle-Hawking wave function for minisuperspace cosmologies \cite{Hartle:1983ai}, then one finds a universe emerge from `nothing'. In this case, it is also common to assume two different arrows of time to find two universes being created from nothing. However, since the two universes are disconnected in the sense that the wave function of the universe vanishes at the initial point, one can only focus on our universe and not worry about the other one. In a similar vein, one might just focus at the surface on which the DeWitt boundary condition is satisfied, and then consider only one of the black holes and not worry about whether there is a bounce picture for this paradigm. As mentioned earlier, the Euclidean core found in the deep quantum regime of black hole models of loop quantum gravity seems to indicate that a simple bounce picture maybe lacking. However, if it is still possible to impose the DeWitt boundary condition on this signature-changing hypersurface, one would have a realization of something like the final-state condition for black holes as proposed in \cite{Horowitz:2003he}. We do not go into details of viability of the final state condition and the DeWitt boundary condition for models of loop quantum gravity which take the dynamical signature-change (as an inescapable implication of the modified gauge structure of spacetime in loop quantum gravity) into account, and leave it for future investigations. However, having a signature-change has been shown to bring cosmological loop models closer to the no-boundary formalism \cite{SigChange_HH}.

Finally, let us end this section by mentioning that there have been previous attempts of naturally finding the DeWitt boundary condition in minisuperspace models of loop quantum cosmology \cite{Bojowald:2001xa}. In that case, it was shown that the difference equation (the analogue of the Wheeler-DeWitt equation in loop quantum cosmology), in the pre-classical (deep quantum) regime, automatically implies a DeWitt-like boundary condition if one is to demand classical behaviour of the wave function at late times. Therefore, it was then claimed that loop quantum cosmology arrives at the DeWitt boundary conditional \textit{dynamically}. Although there have been multiple paradigm-changing developments in the field of loop quantum gravity since the appearance of this paper over two decades ago, it is nice to see that we are also discovering that the DeWitt condition appears naturally in loop quantum black holes as well.

\section{Conclusion}
In this paper, we first reviewed how the DeWitt boundary condition can be applied for the region inside the event horizon. In the anisotropic metric ansatz, one can find a surface where the quantum wave function annihilates, but not at the singularity. However, as the Kantowski-Sachs metric ansatz becomes progressively worse, while approaching the singularity, one has to invoke the BKL conjecture and then the DeWitt boundary condition emerges naturally near the singularity. Our first task was to consider the extension of this to the case of gravitational collapse models.

In order to explore gravitational collapsing cases, we followed the approach of Hajicek and Kiefer to quantize the null shell. One can provide the vanishing boundary condition at the singularity, but we need to consistently embed the shell dynamics into the Penrose diagram. We finally conclude that the Hajicek-Kiefer wave packet explains the shell-antishell pair-annihilation at the singularity due to the destructive interference of the wave function.

From this analysis, we observe that the time-symmetry of the solution is related to the destructive interference which, in turn, leads to the vanishing boundary condition. If we extend this idea to the generic DeWitt boundary condition, one may think that this annihilation condition is the result of the destructive interference between two wave packets, where the existence of not one but two wave packets is justified due to the time-symmetry. If we apply this idea to the space-like singularity of a black hole, then it is not surprising to conclude that the DeWitt boundary condition implies that there exists a black hole and white hole pair-annihilation inside the event horizon. Interestingly, this can be realized in the recent developments of loop quantum gravity inspired black hole models provided we are allowed two arrows of time as has often been done in the past for cosmological models.

In this paper, we reasonably provided a novel interpretation of the loop quantum gravity inspired black holes as well as gravitational collapses of the thin-shells. As a result, we find indications that the DeWitt boundary condition appears a generic and consistent description inside a black hole, even including gravitational collapses, and necessarily requiring quantum gravitational effects. The next interesting question to consider is if the DeWitt boundary condition can be applied as a final-state condition without requiring to invoke the anti-shell (or white-hole) dynamics. This would be particularly suited to the phenomenon of dynamical signature-change in loop quantum gravity which arises as a consequence of non-Riemannian geometries near Planckian regimes. Looking ahead, we would like to answer the most important and physically relevant question of all -- what does such an {annihilation-to-nothing} interpretation of the DeWitt boundary condition mean for the information loss paradox? We leave these interesting questions for future research topics.

\newpage

\section*{Acknowledgment}

DY is supported by the National Research Foundation of Korea (Grant no.:2021R1C1C1008622, 2021R1A4A5031460). SB is supported in part by the NSERC (funding reference \# CITA 490888-16) through a CITA National Fellowship and by a McGill Space Institute fellowship. CYC is supported by the Institute of Physics of Academia Sinica.

\end{document}